\documentclass[preprint,12pt]{aastex}
\usepackage{amsmath}
\usepackage{amssymb}

\begin{document}

\title{Detection of large acoustic energy flux in the solar atmosphere}

\author{
N.~Bello~Gonz\'alez$^{1}$, 
M.~Franz$^{1}$, 
V.~Mart{\'\i}nez Pillet$^{3}$, 
J.A.~Bonet$^{3}$, 
S.\,K.~Solanki$^{2,7}$,
J.C.~del Toro Iniesta$^{4}$, 
W.~Schmidt$^{1}$, 
A.~Gandorfer$^{2}$,
V.~Domingo$^{4}$, 
P.~Barthol$^{2}$, 
T.~Berkefeld$^{1}$, 
M.~Kn\"olker$^{6}$
}

\affil{$^{1}$Kiepenheuer-Institut f\"ur Sonnenphysik, Sch\"oneckstr. 6, 79110 Freiburg, Germany\\
$^{2}$Max Planck Institut f\"ur Sonnensystemforschung, Max-Planck-Strasse 2, 37191 Katlenburg-Lindau, Germany\\
$^{3}$Instituto de Astrof{\'\i}sica de Canarias, Avd. V{\'\i}a L\'actea s/n, La Laguna, Spain\\
$^{4}$Instituto de Astrof{\'\i}sica de Andaluc{\'\i}a (CSIC), Apdo. de Correos 3004, 18080 Granada, Spain\\
$^{5}$Grupo de Astronom\'ia y Ciencias del Espacio, Universidad de Valencia, 46980 Paterna, Spain\\
$^{6}$High Altitude Observatory, National Center for Atmospheric Research\footnote{The National Center for Atmospheric Research is sponsored by the National
Science Foundation.}, Boulder, CO 80307, USA\\
$^{7}$School of Space Research, Kyung Hee University, Yongin, Gyeonggi, 446-701, Korea
}

\email{
nbello@kis.uni-freiburg.de
}

\begin{abstract}

We study the energy flux carried by acoustic waves excited by convective motions at sub-photospheric levels. The analysis of high-resolution spectropolarimetric data taken with IMaX/{\sc Sunrise} provides a total energy flux of $\sim$\,6400--7700\,W\,m$^{-2}$ at a height of $\sim$\,250\,km in the 5.2--10\,mHz range, i.e. at least twice the largest energy flux found in previous works. Our estimate lies within a factor of 2 of the energy flux needed to balance radiative losses from the chromosphere according to \cite{AndersonAthay1989}  and revives interest in acoustic waves for transporting energy to the chromosphere. The acoustic flux is mainly found in the intergranular lanes but also in small rapidly-evolving granules and at the bright borders, forming dark dots and lanes of splitting granules. 
\end{abstract}

\keywords{Sun: photosphere --- Sun: chromosphere --- Sun: oscillations --- techniques: high angular resolution --- techniques: spectroscopic}

\shorttitle{Acoustic waves heating the chromosphere}
\shortauthors{Bello Gonz\'alez et al.}

\section{Introduction}

The long-running debate on the heating of chromospheric layers by acoustic waves is fed by the fact that even the largest measured energy fluxes (found by Bello Gonz\'alez et al.\,2009,\,2010) are nearly an order of magnitude smaller than the amount needed to balance the chromospheric energy losses of 14\,000\,W\,m$^{-2}$  \citep{AndersonAthay1989}. Bello Gonz\'alez et al. found an acoustic energy flux of $\sim$3\,000~W\,m$^{-2}$ at 250\,km and of $\sim$2\,000~W\,m$^{-2}$ at 500\,km from velocity fluctuations measured from narrow-band (FWHM\,=\,1.8\,pm) data at $\sim$\,0{\mbox{$.\!\!^{\prime\prime}$}}4 spatial resolution on the Fe\,{\sc i} lines at 557.6\,nm and 543.4\,nm, formed in mid-photospheric  and low-chromospheric layers, respectively. This is a factor of 4--6 larger than that given by \citet{FossumCarlsson2006}  and \citet{Carlssonetal2007} from intensity fluctuations in continuum bands at 160\,nm (of $\sim$\,1$^{\prime\prime}$ spatial resolution) and in Ca\,{\sc ii}\,H  formed at heights of 430\,km and 200\,km, respectively. We refer the reader to \cite{Belloetal2009} for a detailed introduction to the acoustic-wave heating debate. \\
\indent In this contribution, we present results on acoustic waves from data taken in the Fe\,{\sc i} line at 525.02\,nm with the highest spatial resolution considered so far.

\section{Observations and data analysis}\label{obs}

The data analysed in this study were taken on June 9, 2009 with the IMaX 2D-spetropolarimeter  \citep{MartinezPilletetal2010} onboard {\sc Sunrise} \citep{Bartholetal2010, Solankietal2010}. This study is performed on a time series of $\sim$\,23\,min with 33\,s cadence, taken from 00:36\,UT on in full-polarimetric mode, at  
quiet disc centre. The data consist of narrow (FWHM\,=\,6\,pm) spectropolarimetric filtergrams taken at [$\pm$\,8, $\pm$\,4, +22.7]\,pm around the Fe\,{\sc i}\,525.02\,nm line (Land\'e factor $g$\,=\,3). The capabilities of retrieving the physical parameters with the given spectral sampling have been studied by \cite{Orozcoetal2010}. The data were calibrated according to \cite{MartinezPilletetal2010} (Sect.\,9.2) and 
possess 
a spatial resolution of 0{\mbox{$.\!\!^{\prime\prime}$}}15--0{\mbox{$.\!\!^{\prime\prime}$}}18, achieved by combining the onboard image stabilization system \citep{Gandorferetal2010, Berkefeldetal2010} and phase-diversity techniques \citep{MartinezPilletetal2010}. The 
field of view (FOV) considered is 
44.8$\times$44.8\,arcsec$^2$ (815$\times$815\,pix$^2$). The line-of-sight (LOS) velocities were determined by measuring Doppler shifts of the {\em minimum} position of a Gaussian fitting the 4 spectral points within the Fe\,{\sc i} line plus continuum \citep[][]{MartinezPilletetal2010}.

\subsection{Response functions}\label{RFTFs}

We base our study of wave phenomena on the analysis of fluctuations in the measured LOS velocities.
For a better comprehension of the atmospheric layer where the Fe\,{\sc i} line at 525.02\,nm is sensitive to velocity fluctuations, we calculated velocity response functions ($RF_v$), applying the method described by \cite{Eibeetal2001}. Following \cite{Belloetal2010}, we considered granule (GR) and intergranule (IGR) model atmospheres based on 3D simulations from \cite{Asplundetal2000}. Results are shown in Fig.\,\ref{RFs}a.  In the case of a spectrometer with infinite spectral resolution, the $RF_v$ calculated for the line minimum in both atmospheres are centred around 400\,km height, in agreement with results by \cite{ShchukinaTrujillo2001}.  However, the $RF_v$
are shifted to lower layers and partly broadened when (a) the spectral transmission of the IMaX instrument and (b) the Gaussian-fit used in the determination of the line-minimum velocities, are applied to the emergent intensities in order to mimic the measurements from IMaX observations. For GR, the $RF_v$ is centred at $\sim$\,325\,km and for IGR at $\sim$\,175\,km.\\

\subsection{Atmospheric transmission to velocity amplitudes \label{transmission}}

The broad  extent in height of the $RF_v$ indicates the large atmospheric range where the measured velocities are referred to, i.e. indicates the difficulties in detecting the signal of small-scale velocity fluctuations along the LOS. It is then advantageous to estimate the transmission of the solar atmosphere to (wave) velocity amplitudes, i.e. the ratio of the observed to actual velocities, taking into account the spectral characteristics of our optical system and the velocity determination method.

\begin{figure}[h]
\includegraphics[width=12cm]{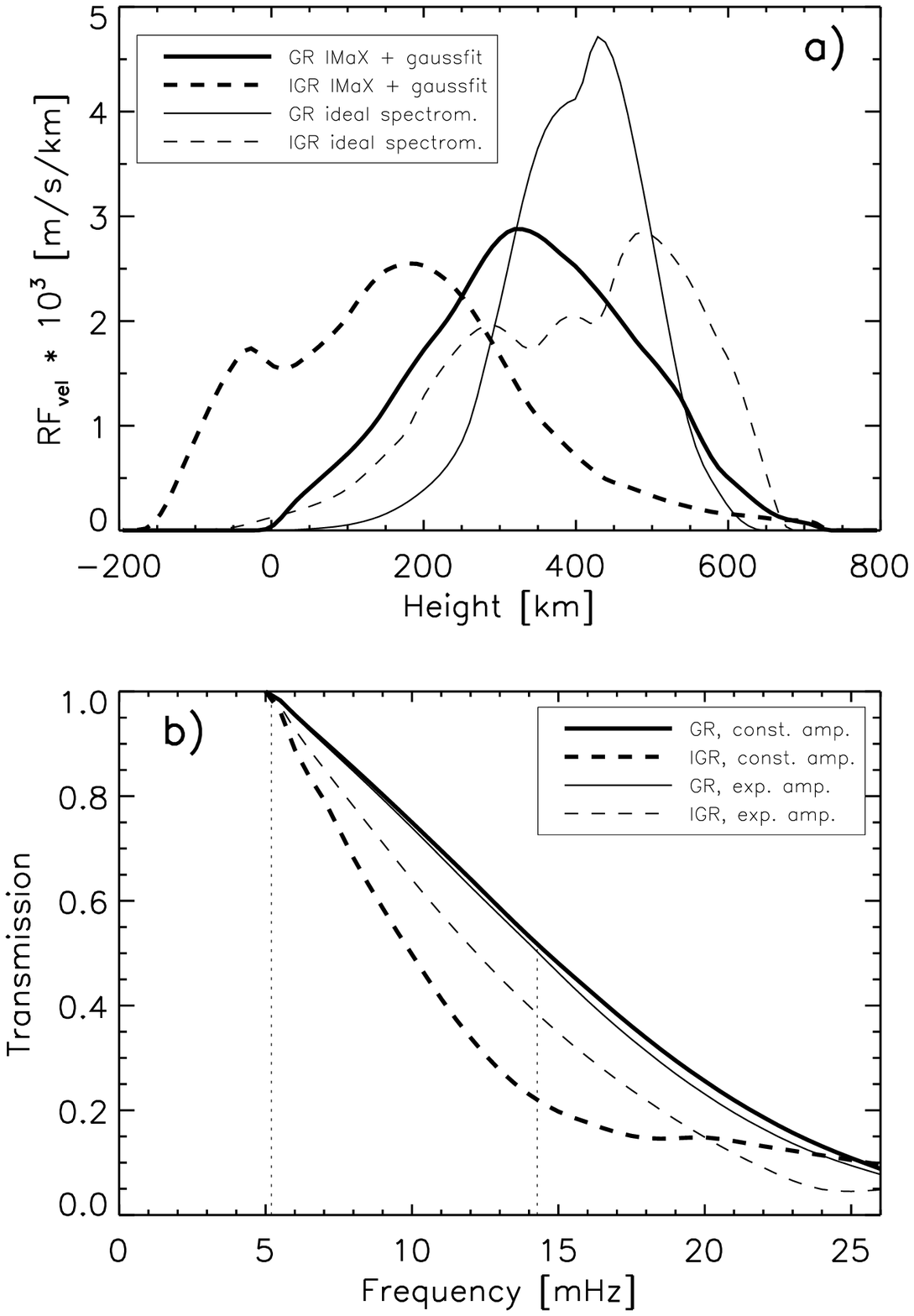}
\caption{{\bf {\em a)}} Response functions for velocity $RF_\upsilon(z)$ at line minimum of Fe\,{\sc i} 525.02\,nm: {\em solid} and {\em dashed}, for granule (GR) and intergranule (IGR) atmospheres, respectively; {\em thick}, after convolution of line profiles with spectrometer function and Gaussian fit, to mimic the IMaX observed profiles.
{\bf {\em b)}} Transmission of solar atmosphere to wave (velocity) amplitudes vs. wave frequency after convolution with the spectral transmission of the IMaX instrument; {\em solid}: GR, {\em dashed} IGR; {\em thick}: waves/ with amplitudes constant in height; {\em thin}: waves with exponentially increasing amplitude. {\em Dotted} lines indicate the frequency (acoustic) range under study, i.e. from the acoustic cut-off frequency, $\nu_{ac}$=5.2\,mHz (U=192\,s)  up to $\nu$=14.3\,mHz (U=70\,s). 
 \label{RFs}}
\end{figure}

Fig\ref{RFs}.

The transmission function is modeled by computing velocities from synthesised IMaX profiles
from a model atmosphere including velocity fluctuations (waves) along the LOS. We consider both the GR and IGR atmospheres following two approaches (i) waves with height-independent amplitude and (ii) with exponentially increasing velocity amplitudes. The latter mimics an acoustic energy flux constant through the stratified atmosphere. We refer to \citet[][]{Belloetal2009,Belloetal2010}  for detailed descriptions on the transmission function calculations and the determination of group velocities derived from the dispersion relation of waves. 

Figure\,\ref{RFs}b depicts the resulting functions. 
They show similar transmissions for the GR atmosphere when considering height-independent or exponentially increasing velocity amplitude. The IGR atmosphere shows a lower transmission for waves with height-independent amplitude. Lower transmission means larger correction in the determination of the energy flux (see Sect.\,\ref{flux}). 

\section{Results}\label{results}

We performed Fourier and wavelet analyses of the 
velocity fluctuations over the time series, 
as in Bello Gonz\'alez et al. (2009, 2010).
The temporal power spectra were calculated at each pixel and averaged over the FOV. 
A separate study on {\em granular} and {\em intergranular} areas has been performed. Granules (GR) are considered to be those regions in the continuum intensity maps where $I_\textrm{cont} / \langle I_\textrm{cont} \rangle > 1$ and intergranules (IGR) where 
$I_\textrm{cont} / \langle I_\textrm{cont} \rangle < 1$. 
Granular dark dots and forming lanes are thus treated as intergranules. 

\subsection{Acoustic energy flux}\label{flux}

The total flux of acoustic energy (per element of area on the Sun) is estimated from the spatially averaged Fourier power $P_\upsilon(\nu_i)$ by
\begin{equation}
F_\mathrm{ac,tot}(\nu_i) = \rho\sum_i P_\upsilon(\nu_i)\cdot \upsilon_\mathrm{group}(\nu_i)/T(\nu_i)\cdot\Delta\nu_i\,,
\label{eq1}
\end{equation}
%
with mass density $\rho$, group velocity $\upsilon_\mathrm{group}$, and frequency intervals $\Delta\nu_i$.  $T(\nu)$ represent the {\em transfer functions}, i.e. the square of the transmission functions 
in Fig.\,\ref{RFs}b. Note that the transfer functions express the smoothing by radiative transfer for small wavelengths and we correct our measurements from this effect. From the GR and IGR models, we adopt the values $\rho_{\textrm{GR}}=5.9\times10^{-5}$~kg\,m$^{-3}$ at 325\,km and  $\rho_{\textrm{IGR}}=10^{-4}$~kg\,m$^{-3}$ at 175\,km. We limit our study to the acoustic frequency range, i.e. from 5.2--15.1\,mHz, 
the latter value being the Nyquist frequency of our observations.

Wavelets are an appropriate tool to retrieve information on the distribution of power in space and time. With this technique we calculated 2D maps of power spectra in period bins of 20\,s from 70\,s up to 190\,s, i.e. the acoustic cutoff period.  We employed the code by \citet{TorrenceCompo1998} with Morlet wavelets and a level of significance of 95\%. An analogous equation to Eq.\,\ref{eq1} is then used in terms of periods $U$ (with  $\Delta U_i$\,=20\,s) to determine the acoustic flux from the wavelet power averaged over time and FOV.

\begin{figure}[h]
\includegraphics{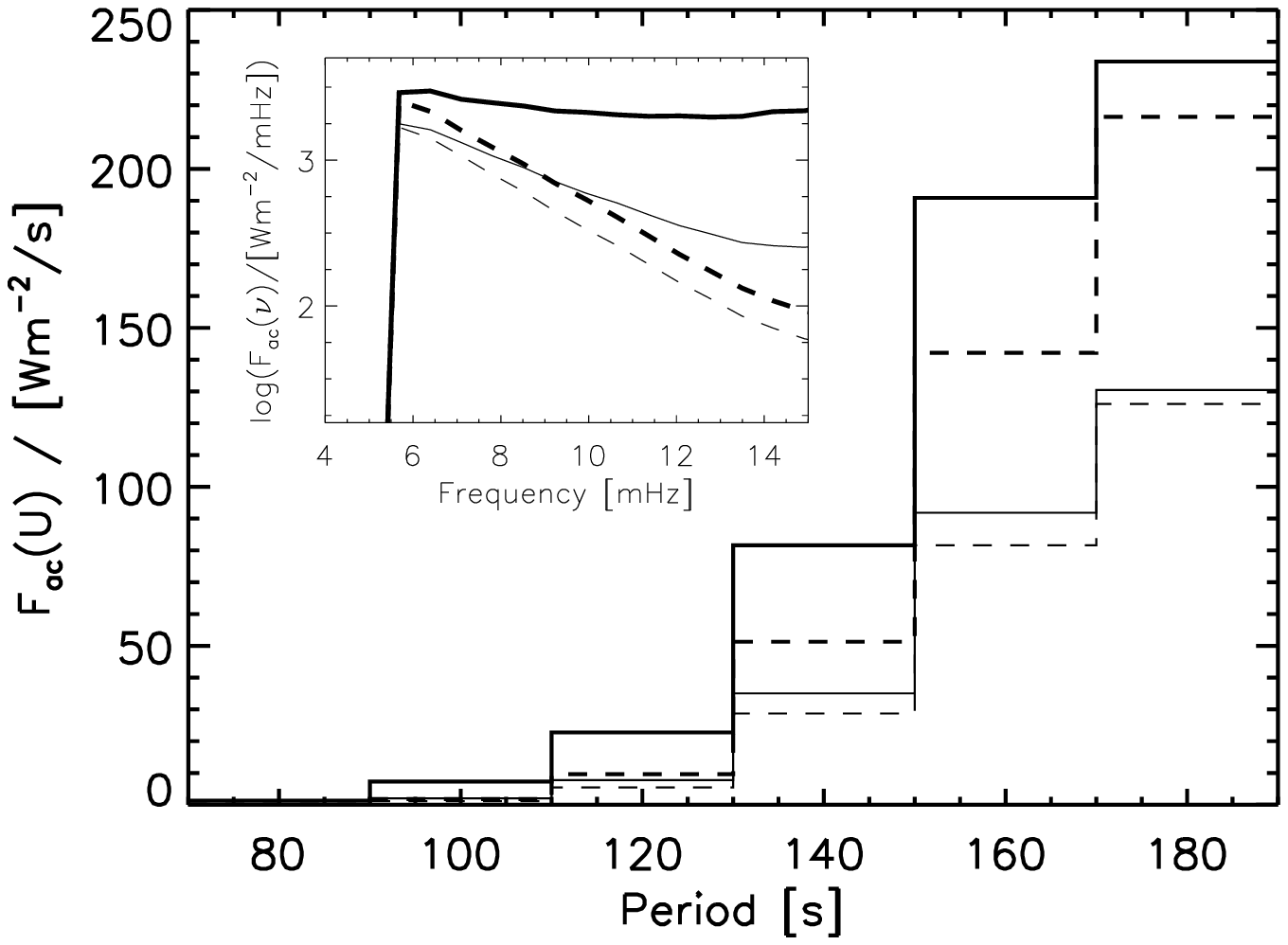}
\caption{Velocity flux spectra from wavelet (vs. period) and Fourier (vs. frequency) analysis. {\em Thin} and {\em thick} represent GR and IGR spectra, before ({\em dashed}) and after ({\em solid}) correction for atmospheric transmission, respectively.  
\label{Flux}}
\end{figure}

Fig\ref{Flux}.

Figure \ref{Flux}  shows averages of velocity flux spectra as 
functions of period (frequency) from the wavelet and Fourier analyses. We differentiate between GR ({\em thin}) and IGR ({\em thick}) flux before ({\em dashed}) and after ({\em solid}) correction of atmospheric transmission, in the case of exponentially increasing amplitudes. This correction is more conservative than that for waves with height-independent amplitude (see Fig.\,\ref{RFs}b), therefore it will provide us with lower-limit corrected estimates. For the differentiation we obtain the power above GRs and IGRs from the wavelet analysis (cf. Fig.~\ref{pow_cont}). The resulting weights are transferred to the power in the Fourier analysis \citep[see][]{Belloetal2010}.

Before correction, we find in both the Fourier and wavelet analyses, an IGR flux larger than the GR flux by a factor 1.7. The correction increases the flux in the IGR more strongly, as expected from the transmission functions (Fig.\,\ref{RFs}b). After the correction, the ratio between IGR and GR fluxes amounts to 1.8--2.0.

Table\ref{table1}.
\begin{table}
\caption{Total acoustic energy fluxes, $F_{ac, tot} \ [W\,m^{-2}]$\label{table1}}
\begin{tabular}{lcccc}
\tableline
\tableline
wavelet:  &  & uncorr & corrected  \\
\tableline
GR  &   & 4430 &  5000--4970\\
IGR  &     & 6400 &  7725--8630   \\
average  &  &  5495  &  6470--6945  \\
\tableline
5.2--10\,mHz  &    & 5455 &  6360--6765  \\
10--14.3\,mHz    &  & 40 &  110--180  \\
\tableline
\tableline
Fourier: &  & uncorr & corrected \\
\tableline
GR  &   & 4930 &   7000--6865   \\
IGR  &    &7540 &  13075--20165   \\
average  & &    6340 &  10280--14050  \\
\tableline
5.2--10\,mHz   & &    5630 &  7720--8900   \\
10--14.3\,mHz  & &    710 &  2560--5150   \\
\tableline
\tableline
\end{tabular}
\tablecomments{{\em Left} and {\em right} values in column {\em corrected} refer to the exponentially increasing and height-independent velocity amplitude approaches, respectively (Sect.\,\ref{transmission}).}

\end{table}

\

The results on the total acoustic flux summed over all frequencies are collected in Table\,\ref{table1}. The 'average' values refer to the flux over the FOV filled by GR over 46\% of the area and by IGR over the remaining 54\%. Results from the wavelet and Fourier analyses agree for the most part at low frequencies, i.e. for periods $\ge$\,100\,s. They amount to $\sim$\,5500\,W\,m$^{-2}$ before correction and to $\sim$\,6400--7700\,W\,m$^{-2}$ after the conservative correction. Yet, a large difference appears at high frequencies, where the treatment of the Fourier and wavelet analysis is different: 
(1) The wavelet approach is more restrictive in terms of noise when using a statistical significance level of 95\%, therefore the retrieved values at these frequencies are low, 
(2) the wavelets were applied in period bins which reduces the spectral sampled points at high frequencies to just the bin at 80\,s and half of the bin at 100\,s,
(3) the values at high frequencies are strongly corrected, i.e. the originally larger values in the Fourier analysis spectra are highly increased.

We are confident in the correction of the flux spectra for frequencies $\le$\,10\,mHz. The large correction of flux for frequencies $>$\,10\,mHz above IGRs from the Fourier analysis, see the inset in Fig.~\ref{Flux}, turns the flux spectrum upward. This also was noted by \cite{Flecketal2008} who point out the difficulty in determining the noise level. In addition, aliasing from power at frequencies above our Nyquist frequency may play a role. With these caveats, we consider the large fluxes at high frequencies as upper limits. The high frequency waves are most prone to radiative damping. Their amplitudes will thus be strongly reduced at the base of the chromosphere.

In any case, even before correction, the directly measured energy fluxes are larger than those found in any previous observational study of acoustic waves. The nature of these results and their consequences are discussed in Sect.\,\ref{Discussion}.

\subsection{Spatial distribution of the acoustic flux \label{powerocc}}

\begin{figure}[h]
\includegraphics[width =12cm]{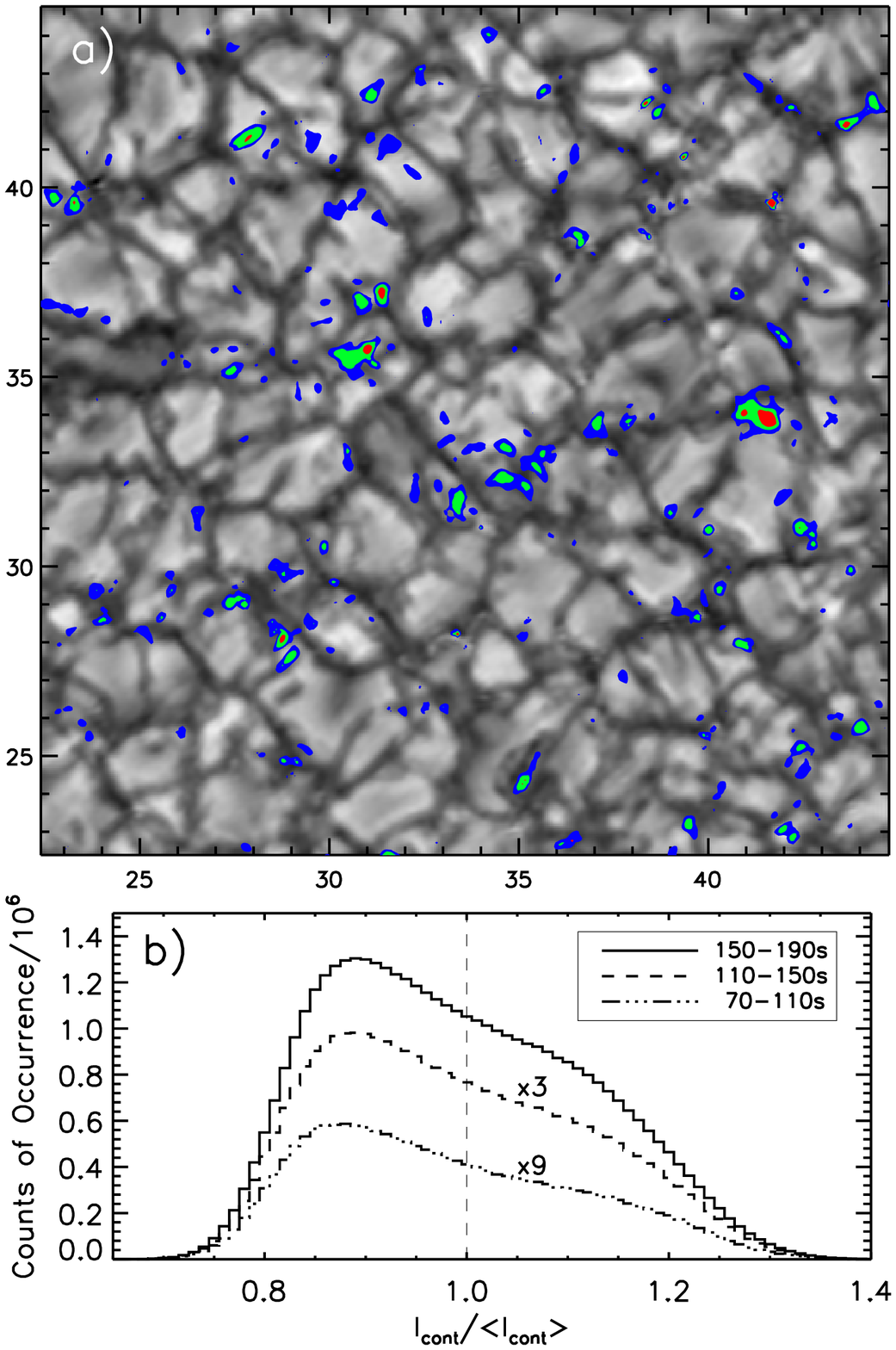}
\caption{{\bf {\em a)}} Snapshot on acoustic power averaged over the period range $[$150--190$]$\,s, overlaid to a continuum image. The area corresponds to the upper-right quadrant of the full FOV; filled contours represent 12\% (blue), 20\% (green) and 32\% (red) of the maximum power value over the sequence; tickmarks in arcsec. The underlying animation is available at URL. {\bf {\em b)}} Occurrence of wavelet power of velocities above 1$\%$ of maximum power vs. normalised continuum intensities $I_\textrm{cont}/\left\langle I_\textrm{cont}\right\rangle$ for reconstructed data. Curves for $[$110-150$]$\,s and $[$70-110$]$\,s period bins were amplified by a factor 3 and 9, respectively, for better visualisation.
\label{pow_cont}}
\end{figure}

The wavelet analysis provides maps of wave power for each time step. With this technique applied to the high-spatial resolution IMaX data, we can study in detail the distribution of power (flux) over the observed FOV and its evolution with time. Figure\,\ref{pow_cont} shows an example of different levels of acoustic power  over the period range [150--190]\,s overlaid on continuum images, and an animation on the time evolution of such power patches 
is available at URL.
 
Fig\ref{pow_cont}.

The time lag corresponding approximately to the travel time from the bottom of the photosphere to 250\,km height is about 42\,s. The cadence of the time series is 33\,s, therefore no delay between intensity and power maps has been taken into account. 

The wave power, far from being homogenously distributed, appears
intermittently in space and time as already found by, e.g., \cite{Rimmeleetal1995}, \cite{Wunnenbergetal2002} and \cite{Belloetal2009,Belloetal2010}. Histograms of the acoustic power occurrence over intensity
represented in Fig.\,\ref{pow_cont} show that most of the power, at all frequencies, is associated with {\em dark} structures, i.e. structures with intensities $I_\textrm{cont}/ \langle I_\textrm{cont} \rangle < 1$. We have referred to these structures as {\em intergranules}, but close inspection of the power maps overlaid on the continuum intensities (see animation available at URL, attached to this manuscript) reveal that significant power is found in forming granular (dark)  dots and lanes. A second, less prominent component or shoulder is seen in the histograms  at $I_\textrm{cont}/ \langle I_\textrm{cont}\rangle \sim 1.1$. It has not been reported 
before, we see it thanks to the high contrast of the IMaX images. This power is found in small rapidly evolving granules, and in large granules (1) prior to their splitting and (2) at their bright borders. Yet not all splitting granules and bright borders show a wave power component.

\subsection{Energetic fast-fluctuating features\label{events}}

Studying the evolution of acoustic power reveals patches of intense signal. They appear intermittently and randomly distributed, covering areas of typically 1\,arcsec$^{2}$ and with power at all periods lasting for about 5\,min. Such strong signals are found to be generated by rapid fluctuations in velocities up to $\sim$\,4\,km\,s$^{-1}$, that are of a diverse nature: 
(1) sudden (within 2--3\,min) up-down-up flows, 
(2) sudden down-up-down flows, 
(3) some associated to magnetic field concentrations, 
(4) some with no significant magnetism, 
(5) some of the latter related to imploding granular areas, etc. 

In a first analysis, we have identified 32 of these events within the 23\,min series. Masking these events and calculating the acoustic flux from their wave power, we find that they carry 
20--23\,kW\,m$^{-2}$ (estimated values before and after correction, respectively). Yet, their small area coverage, 0.1\% of the FOV, implies that their global contribution to the total flux is not significant. 

Nevertheless, the identification of these energetic features at small scales from their strong power signature has opened a new field of study and we refer to a forthcoming contribution on the nature of these events.

\section{Discussion and conclusions}\label{Discussion}

In this study we have taken advantage of the high quality data of the IMaX/{\sc Sunrise} instrument. The high spatial resolution provided new exciting findings, which (a) confirm that acoustic waves play an important role in the energy transport through the quiet solar atmosphere, (b) shed new light on the spatial and temporal distribution of propagating wave phenomena and (c) reveal the existence of very energetic processes occurring at small scales.

Table\,\ref{table2} summarises the values of acoustic energy flux measured by different authors in the last years. Averaged values from the present study before and after correction of the transmission of the solar atmosphere to the velocity wave amplitudes are also listed for comparison. 

Table\ref{table2}.
\begin{table}
\caption{Summary of observed total acoustic flux $F_{ac, tot}$\,$[Wm^{-2}]$ \label{table2}.}
\begin{tabular}{lcc}
\tableline
\tableline
{\bf method / data} & uncorr. & corrected \\
\tableline
\tableline
\tableline
wavelet: binned (2$\times$2) &   3975 &  4600--4900   \\
\tableline
wavelet:  binned (1$\times$1)  &  5495  &  6470--6945  \\
\tableline
Fourier: binned (1$\times$1)  &   6340 & 10280--14050   \\
\tableline
\tableline
{\bf other work} \\
\tableline
\tableline
Bello Gonz\'alez et al. (2010) \\
$h=500-600\,km$, {0\mbox{$.\!\!^{\prime\prime}$}}4 & -- &1700--2000 \\
\tableline
Bello Gonz\'alez et al. (2009) \\
 $h=250\,km$, {0\mbox{$.\!\!^{\prime\prime}$}}4 &  2500 &  3000--3650\\
\tableline
Straus et al. (2008)\\
 $h=250\,km$,  $\sim$\,{0\mbox{$.\!\!^{\prime\prime}$}}4 & -- &1400 \\
 $h=500\,km$,  $\sim$\,{0\mbox{$.\!\!^{\prime\prime}$}}5 & -- &1000 \\
\tableline
Carlsson et al (2007)\\
$h=200\,km$, $\sim$\,{0\mbox{$.\!\!^{\prime\prime}$}}22 & --  &800 \\
\tableline
Fossum \& Carlsson (2006)\\
 $h=430\,km$,   $\sim$\,1$^{\prime\prime}$ &  -- &510\\
\tableline
Wunnenberg et al. (2002) \\
$h=600\,km$,  $\sim$\,{0\mbox{$.\!\!^{\prime\prime}$}}5 & -- & 900 \\
\tableline

\tableline
\tableline
\end{tabular}
\tablecomments{Notation in column {\em corrected} as in Table \ref{table1}.}

\end{table}

We note that the use of transmission functions for  correcting velocity amplitudes at {\em short periods} is controversial. The results from 3D hydrodynamic simulations by \cite{Flecketal2010} suggest that the observed high frequency Doppler shifts are caused by rapid variations of the 
contribution functions in an atmosphere with high velocity gradients. These authors conclude that one should re-evaluate the cause of Doppler shifts with periods $\le70$~s, i.e. periods below the Nyquist period of our observations.

We restrict the following discussion to the flux values from the wavelet analysis, which we consider to be {\em lower limits}.
 The average acoustic flux at 250\,km is 
$F_{ac,tot}\,\sim\,6\,500\,W\,m^{-2}$, after a 
conservative\footnote{{\em conservative} stands for the less restrictive correction when calculating transfer functions for exponentially increasing velocity amplitudes, Sect.\,\ref{RFTFs}.} correction. It is mainly found at low frequencies (periods $\ge$\,100\,s). Yet, the uncorrected flux already amounts to \,5\,500\,W\,m$^{-2}$. These large values were never reported before, probably due to limited spatial resolution. \cite{WedemeyerBoehmetal2007} pointed out from 3D simulations that a significant amount of acoustic power is lost when degrading the spatial resolution. This effect can be seen when applying to our data a 2$\times$2 pixel binning. The signal is reduced by a factor of 1.4. 
We note that \cite{Carlssonetal2007}, analysing HINODE data, found a lower power reduction upon binning.
Also compare the resulting acoustic flux with that found by \cite{Belloetal2009} at the same height. There the spatial resolution 
is at least a factor 2 less, although the spectral sampling is 
better. After transmission correction, the IMaX data 
provide fluxes larger by a factor of 2. We conclude that much of the power in the quiet Sun occurs at small scales and high spatial resolution is mandatory to detect it.\\

Simulations by \cite{Strausetal2008} give an estimate of 30\,$\%$ for energy losses between 250\,km up to 500\,km due to radiative damping. 
Applying the same correction, the averaged acoustic flux we observe at 250 km would then be reduced to 4500\,W\,m$^{-2}$ at the bottom of the chromosphere.
 On the one hand, the flux reduction 
between 250\,km and 500\,km may be larger because of the very small spatial scales in the present data set. On the other hand, the estimated flux at 250\,km is only a lower-limit value since we have considered: 
(i) wavelet power, 
(ii) the conservative correction. 
This acoustic flux is comparable to the radiative losses of 4600\,W\,m$^{-2}$  quantified from the standard semi-empirical hydrostatic model of a quiet average chromosphere by \cite{Vernazzaetal1981}. Chromospheric models calculated including radiative losses from Fe\,{\sc ii} lines, arrive at energy needs 
of 14\,000\,W\,m$^{-2}$ \citep{AndersonAthay1989}. In our wavelet analysis, the contribution from high-frequency ($>$\,10\,mHz) waves is missing, but likely present in the solar atmosphere as suggested, e.g. the Fourier analysis.

Let us recall the main findings of this first study on acoustic waves from IMaX/{\sc Sunrise} observations at disc centre:\\

1) The improvement in at least a factor of 2 in spatial resolution and the image quality of IMaX/{\sc Sunrise} data
have revealed a total acoustic power larger by at least a factor of 2 than the largest power observed before. In view of the results obtained on the total acoustic power, we conclude that, at photospheric level, acoustic waves with periods $>$\,100\,s carry at least half of the flux needed to balance the observed radiative energy losses of the quiet chromosphere according to \cite{AndersonAthay1989}. 
The possibility that the quiet solar chromosphere is supplied with acoustic wave energy to cover its radiative losses, as suggested many years ago by L. Biermann and M. Schwarzschild and later on pursued intensively by P. Ulmschneider, 
needs to be revived and intensively studied, probing which fraction of the transported energy reaches the chromosphere.\\

2) The main contribution to the acoustic energy flux is found in intergranules, as is extensively known. In addition, forming dark dots and lanes above splitting granules as well as small fast-evolving granules and the brighter borders of large granules are found to be sources of propagating waves. These are regions of strong dynamics (turbulence) as pointed out in the studies on the same IMaX data series by \cite{Steineretal2010},  and \cite{Rothetal2010}. \\

3) Thanks to the high quality of the IMaX data, it has been possible to detect intermittent sources of strong energy fluxes at all frequencies in the acoustic domain. A preliminary analysis of their nature shows that they are generated by rapid (2--3\,min) fluctuations in their LOS velocity occurring at small scales, i.e. typically of 0{\mbox{$.\!\!^{\prime\prime}$}}3\,$\times$\,0{\mbox{$.\!\!^{\prime\prime}$}}3.. They show a diverse origin, in some cases they are related to small magnetic field concentrations, in others to imploding granules, etc. 
We refer to a forthcoming contribution on a detailed analysis of these events.

\begin{acknowledgements}
NBG wants to thank F. Kneer for his invaluable help and interest in this work. 
NBG also wants to 
thanks R. Schlichenmaier and B. Louvel for 
helpful 
suggestions on the manuscript.
   The German contribution to $Sunrise$ is funded by the Bundesministerium
   f\"{u}r Wirtschaft und Technologie through Deutsches Zentrum f\"{u}r Luft-
   und Raumfahrt e.V. (DLR), Grant No. 50~OU~0401, and by the Innovationsfond of
   the President of the Max Planck Society (MPG). The Spanish contribution 
   was funded by the Spanish MICINN under projects ESP2006-13030-C06 and
   AYA2009-14105-C06 (including European FEDER funds). The HAO contribution was
   partly funded through NASA grant number NNX08AH38G. 
Support by the WCU grant No.\,R31--10016 by the Korean Ministry of Education, Science and Technology is acknowledged.
\end{acknowledgements}

\bibliographystyle{apj}


\end{document}